
\magnification=1100
\overfullrule0pt

\input amssym.def



\newsymbol\ltimes 226E 
\newsymbol\rtimes 226F 


\font\smallcaps=cmcsc10
\font\titlefont=cmr10 scaled \magstep1

\font\sectionfont=cmbx10
\font\tinyrm=cmr10 at 8pt


\newcount\sectno
\newcount\subsectno
\newcount\resultno

\def\section #1. #2\par{
\sectno=#1
\resultno=0
\bigskip
\centerline{\sectionfont #1.  #2} }

\def\subsection #1\par{\bigskip\noindent{\it  #1} \medbreak}


\def\prop{ \global\advance\resultno by 1
\bigskip\noindent{\bf Proposition \the\sectno.\the\resultno. }\sl}
\def\lemma{ \global\advance\resultno by 1
\bigskip\noindent{\bf Lemma \the\sectno.\the\resultno. }
\sl}
\def\remark{ \global\advance\resultno by 1
\bigskip\noindent{\bf Remark \the\sectno.\the\resultno. }}
\def\example{ \global\advance\resultno by 1
\bigskip\noindent{\bf Example \the\sectno.\the\resultno. }}
\def\cor{ \global\advance\resultno by 1
\bigskip\noindent{\bf Corollary \the\sectno.\the\resultno. }\sl}
\def\thm{ \global\advance\resultno by 1
\bigskip\noindent{\bf Theorem \the\sectno.\the\resultno. }\sl}
\def\defn{ \global\advance\resultno by 1
\bigskip\noindent{\it Definition \the\sectno.\the\resultno. }\slrm}


\def\formula{\global\advance\resultno by 1
\eqno{(\the\sectno.\the\resultno)}}
\def\formulano{\global\advance\resultno by 1 (\the\sectno.\the\resultno)}
\def\tableno{\global\advance\resultno by 1
\the\sectno.\the\resultno. }
\def\lformula{\global\advance\resultno by 1
\leqno(\the\sectno.\the\resultno)}


\def\monthname {\ifcase\month\or January\or February\or March\or April\or
May\or June\or
July\or August\or September\or October\or November\or December\fi}

\newcount\mins  \newcount\hours  \hours=\time \mins=\time
\def\now{\divide\hours by60 \multiply\hours by60 \advance\mins by-\hours
     \divide\hours by60         
     \ifnum\hours>12 \advance\hours by-12
       \number\hours:\ifnum\mins<10 0\fi\number\mins\ P.M.\else
       \number\hours:\ifnum\mins<10 0\fi\number\mins\ A.M.\fi}


\nopagenumbers
\def\runningtitle{\smallcaps black hole entropy}
\headline={\ifnum\pageno>1\eoheadline\else\firstheadline\fi}
\def\names{\smallcaps b. ram}
\def\firstheadline{\noindent Preliminary Draft \hfill  \today}
\def\firstheadline{}
\def\eoheadline{\ifodd\pageno\oddheadline\else\evenheadline\fi}
\def\oddheadline{\tenrm\hfil\runningtitle\hfil\folio}
\def\evenheadline{\tenrm \folio\hfil{\names}\hfil}

\vphantom{$ $}  
\vskip.75truein

\centerline{\titlefont The mass quantum and black hole entropy II}
\bigskip
\bigskip
\bigskip
\centerline{\titlefont B. Ram}
\bigskip
\bigskip

\centerline{Physics Department,  New Mexico State University}
\centerline{Las Cruces, NM 88003 USA${}^{\dagger}$}
\medskip
\centerline{and}
\medskip
\centerline{Umrao Institute of Fundamental Research}
\centerline{A2/214 Janak Puri, New Delhi, 110058, India}


\footnote{}{\tinyrm ${}^\dagger$ Correspondence address.}

\bigskip
\bigskip
\bigskip
\bigskip

\hrule
\bigskip\bigskip

\noindent{\bf Abstract}

\bigskip
In [Phys. Lett. A 265 (2000) 1] a new method was given
which naturally led to a quantum of mass equal to twice
the Planck mass.  In the present note which, for convenience,
we write formally as a continuation of that paper, we show that
with spin one of the mass quantum, the physical entropy of a rotating black
hole is also given by the Bekenstein-Hawking formula.


\bigskip\bigskip
\hrule
\bigskip\bigskip

The interest in this theory [1] lies in the facts:
(1) that it combines elements of general relativity with 
elements of quantum mechanics in an unprecedented way; (2) that it 
gives the correct physical result for the entropy of a black
hole, namely $A/4$, which has virtually
been elevated to the status of a cardinal truth.

One may observe that the quantum eigenvalue equation (8)
can be obtained by simply setting $E=\ell=0$ in Eq.\ (5).  This is
equivalent to putting the classical constants of integration $E$ and $L$ equal to 
zero in Eq.\ (1) and then applying the Schr\"odinger prescription.

Let us apply this recipe to the classical time like Kerr
geodesic [2]:
$${1\over 2}\dot r^2-{m\over r}+{1\over 2}(1-E^2)(1+a^2/r^2)
+{L^2\over 2r^2} -{m\over r^3}(L-aE)^2=0,
\eqno{(28)}$$
where $a$ is the angular momentum per unit mass of the source [3].
Putting $E=L=0$ in (28), followed by the Schr\"odinger prescription,
gives
$$\left( -{1\over 2r^2}{d\over dr}\right.\left(r^2{d\over dr}\right)
\left.
+{j(j+1)\over 2r^2} -{m\over r}\right) R = -{1\over 2} R,
\eqno{(29)}
$$
or
$$
\left[ -{1\over 2}\right.\left(
{d^2\over dr^2}-{j(j+1)\over r^2}\right)\left. - {\mu/4\over r}\right] U
=-{1\over 2}U,
\eqno{(30)}
$$
with $U=rR(r)$ and $m=\mu/4$, as the quantum equation for the 
{\it pure} mass $\mu$.  Eq.\ (30) is Eq.\ (11) with
$$\ell'=j,\quad N'=3, \quad B=1/2, \quad \alpha=\mu/4;$$
and hence it represents a {\it four dimensional harmonic oscillator}
with $\omega=2$, and 
$$\mu_{n,j} = 2(n+j+1)\omega,\qquad n=0,1,2,\ldots
\eqno{(31)}
$$
and the values of $j$ determined as follows:

In modern language Eq.\ (31) says that the $n$th mass $(\mu)$ states are
occupied by $n$ {\it pairs} of mass quanta and $2j$ is the 
spin angular momentum of the $n$ pairs.

As $n=1$ states have only one pair, $2j$ is the spin (for 
brevity) of the pair.  Let us designate by $j_q$ the {\it spin}
of the mass quantum, which can have only two values along the $z$-axis [3],
$+j_q$ (up) and $-j_q$ (down).  In principle $j_q$ can have any 
positive integral value.  We choose $j_q=1$, for
simplicity and because this value is in accord with the quantum
version of the censorship hypothesis [4], namely
$$\sqrt{ j_q(j_q+1) } \le 2 = \hbox{ mass of the quantum}.$$
Now the pair can be formed in only two ways: 
(i) when the spins of the two quanta are anti-aligned; and 
(ii) when the spins of the two quanta are aligned {\it up}.
In case (i) the spin of the pair is zero, giving $2j=0$, or $j=0$;
in case (ii) the spin of the pair is two, giving $2j=2$, or $j=1$.
(The value $j=-1$ is not allowed because in Eq.\ (30) the quantum 
number $j$ is a positive integer or zero.)

Next consider the $n=2$ states.  They have two pairs.  These
can be formed in three ways.  (1) Both pairs have spin zero,
giving $j=0$; (2) one pair has spin zero and the other has spin 2,
giving $j=1$; (3) each pair has spin 2,
giving $2j=4$, or $j=2$.

Continuing the process leads to the values $j=0,1,2,\ldots, n$
for the $n$th mass states which contain $n$ pairs.

Clearly then Eq.\ (31) can be rewritten as
$$\mu_{n'_{n,j}}=2(n'_{n,j}+1)\omega,
\qquad n'_{n,j} = 0,1,2,\ldots
\eqno{(32)}
$$

It is to be noted that Eq.\ (32) is Eq.\ ($9'$)
when every one of the $n$ pairs has spin zero.

It follows straightforwardly from Eq.\ (32) that the physical
entropy of a non-extremal Kerr black hole of mass $M$ and
angular momentum $a$ should be exactly given by the relation
(27) with $A=16\pi M_{ir}^2$, provided the physical temperature 
of the black hole is defined by
$$T = {\kappa\over 2\pi (1-a^2/M^2)^{1/2}} = {\kappa'\over 2\pi}.
\eqno{(33)}
$$

In conclusion we may say that the two parameters that characterize
the Kerr solution of classical general relativity are fused by
quantum mechanics into a single concept of a quantum of mass {\it twice}
the Planck mass and spin angular momentum {\it one}.

\bigskip\bigskip
\noindent
{\bf Acknowledgements}
\smallskip
The author thanks A.\ Ram, N.\ Banerjee, R.S.\ Bhalerao, and A.\ Bussian for 
useful conversations, V.\ Singh for affirming comments, N.\ Ram for critical 
reading of the manuscript and making concrete suggestions, and the Tata 
Institute for a pleasant stay.

\bigskip\bigskip

\centerline{\smallcaps References}

\medskip
\item{[1]} {\smallcaps B.\ Ram}, 
Phys. Lett. A 265 (2000) 1 [gr-qc/9908036].

\medskip
\item{[2]} {\smallcaps S.\ Chandrasekhar}, 
{\sl The Mathematical Theory of Black Holes},
Oxford University Press, 1983.

\medskip
\item{[3]} {\smallcaps R.P.\ Kerr},
Phys. Rev. Lett. 11 (1963) 237.

\medskip
\item{[4]} {\smallcaps B.\ Carter},
in {\sl Black Holes}, Editors C.\ DeWitt and B.S.\ DeWitt,
Gordon and Breach, 1972.

\vfill\eject
\end